\documentclass{aa}
\usepackage{psfig}

\begin{document}

   \thesaurus{06         
              (08.02.3;  
               08.14.2;  
               08.09.02 4U~1812--12;  
               13.25.1;  
               13.25.5)}  
   \title{Observations of Eddington-limited type-I X-ray bursts from 4U~1812--12}

   \author{ M. Cocchi \inst{1}, A. Bazzano \inst{1}, L.Natalucci \inst{1}, P. Ubertini \inst{1},
            J. Heise \inst{2}, E. Kuulkers \inst{2,3}, J.M. Muller \inst{2,4} and J.J.M. in 't Zand \inst{2}
          }
   \offprints{M. Cocchi, wood@ias.rm.cnr.it}

   \institute{$^{1}$ Istituto di Astrofisica Spaziale ({\em IAS/CNR}), via Fosso del Cavaliere 100, 00133 Roma, Italy \\
              $^{2}$ Space Research Organization Netherlands {\em (SRON)}, Sorbonnelaan 2, 3584 TA, Utrecht, The Netherlands \\
              $^{3}$ Astronomical Institute, Utrecht University, P.O. Box 80000, 3507 TA, Utrecht, The Netherlands \\
              $^{4}$ {\em BeppoSAX} Science Data Centre, Nuova Telespazio, via Corcolle 19, 00131 Roma, Italy
             }

   \date{Received ?; accepted ?}

   \authorrunning{Cocchi et al.}
   \maketitle

   \begin{abstract}
During more than 3 years (August 1996--October 1999) monitoring of a $40\degr \times 40\degr$ 
sky region around the Galactic Centre by the Wide Field Cameras
on board {\em BeppoSAX}, a total of 8 type-I bursts have been detected from a sky position 
consistent with that of 4U~1812$-$12, a likely neutron-star low-mass X-ray 
binary.  We present the results of a detailed study of the bursts of 4U~1812$-$12, about 15
years after the last reported observations of X-ray bursts from this source (\cite{Mura83}).
Clear evidence for photospheric radius expansion due to Eddington-limited burst 
luminosity is present in most of the observed events, allowing an accurate 
estimate of the source distance ($\sim 4$ kpc) and its burst parameters.

   \end{abstract}


\section{Introduction}

Since its first {\em Uhuru} detections (\cite{Form76,Form78}), 4U~1812$-$12 was observed by several  
satellite X-ray experiments: {\em OSO~7} (1M~1812-121, \cite{Mark79}), 
{\em Ariel~V} (3A~1812-121, \cite{Warw81}), {\em HEAO}~1 (1H~1815$-$121, \cite{Wood84}), 
and {\em EXOSAT} (GPS~1812-120, \cite{Warw88}).
From these observations, it is clear that 4U~1812$-$12 is a persistent, though
variable, source.  {\em Uhuru} found a 2--10 keV maximum intensity of 
$\sim 5\times 10^{-10}~{\rm erg~cm}^{-2}{\rm s}^{-1}$ and a variability of a factor of at least 2
(\cite{Form78}).  Similar variability characteristics were observed by {\em Ariel~V}, as the source
varied in the range $\sim 3-6\times 10^{-10}~{\rm erg~cm}^{-2}{\rm s}^{-1}$ in the same energy band 
(\cite{Warw81}), while lower intensities of $\sim 2\times 10^{-10}~{\rm erg~cm}^{-2}{\rm s}^{-1}$ 
and $\sim 3\times 10^{-10}~{\rm erg~cm}^{-2}{\rm s}^{-1}$ were measured by {\em HEAO}~1 
(2--10 keV, \cite{Wood84}) and {\em EXOSAT} (2--6 keV, \cite{Warw88}).
The 3--10 keV source spectrum as obtained by the {\em EXOSAT} GSPC was best fitted by a power law 
(\cite{Gott95}).
4U~1812$-$12 is being monitored by {\em RXTE}-ASM since February 1996 
\footnote{the ASM measurements are publicly available at URL {\tt http://www.space.mit.edu/XTE}}, 
confirming its previously reported characteristics.  
The source is always detected, with an average 2--10 keV flux of $\sim 3.8\times 10^{-10}~{\rm erg~cm}^{-2}{\rm s}^{-1}$ 
($\sim 20$ mCrab) and a variability of a factor $\la 3$ on $\sim 1$ week time scale.

Three X-ray bursts were detected from this source in 1982 by {\em Hakucho} (\cite{Mura83}).
Two of the events showed clear evidence for photospheric radius expansion, and 
reached a maximum 1--22 keV intensity of $\sim 1.7\times 10^{-7}~{\rm erg~cm}^{-2}{\rm s}^{-1}$.
The burst spectra were consistent with a $\sim 2.5$ keV blackbody emission, and showed evidence for softening 
during the exponential decay ({\em e}-folding time $\tau \sim 20$ s). 
This indicated the bursts to be type-I, i.e. thermonuclear flashes
originating on the hot surface of a neutron star, and the source to be likely located in a 
low-mass X-ray binary.
During the observation no persistent emission was detected above 20 Uhuru flux units 
($\sim 5\times 10^{-10}~{\rm erg~cm}^{-2}{\rm s}^{-1}$ in 2-10 keV). Anyway this is not 
in disagreement with the identification of the {\em Hakucho} burster with the persistent source 4U~1812$-$12.
\cite{Mura83} also proposed the association of the burst source with the transient Ser~X-2, observed 
once in 1965 (\cite{Frie67}).
More recently, a single-peaked burst from 4U~1812$-$12 was observed by the {\em BeppoSAX}-WFC 
instrument on 1997 Mar 12.2209 UT (\cite{Burd97}).  The event had a peak intensity of 1.2 and 0.6 Crab in the 
1.5--10 keV and 10$-$26 keV band respectively, and a decay time of $\sim 20$ s.

4U~1812$-$12 is classified as an atoll source, which is
common among the type-I X-ray bursters, and shows band-limited noise and a $\sim 0.8$ Hz QPO (\cite{Wijn99}).

In this paper we investigate the burst characteristics of 4U~1812$-$12, about 15 years after the 
{\em Hakucho} observations of its type-I bursts. The event observed by Burderi et al. (1997) is also re-analysed.
Through a homogeneus sample of Eddington-limited type-I X-ray bursts, we accurately estimate the source 
distance and test the reliabilty of near-Eddington bursts as a standard candle.
In the next section we briefly introduce the Wide Field Cameras telescopes and report on the observations 
of 4U~1812$-$12.  Time-resolved spectroscopy
of the bursts is presented in Section 3, while the scientific implications of our results are 
discussed in Section 4.

    \begin{figure}[hbt]
      \psfig{figure=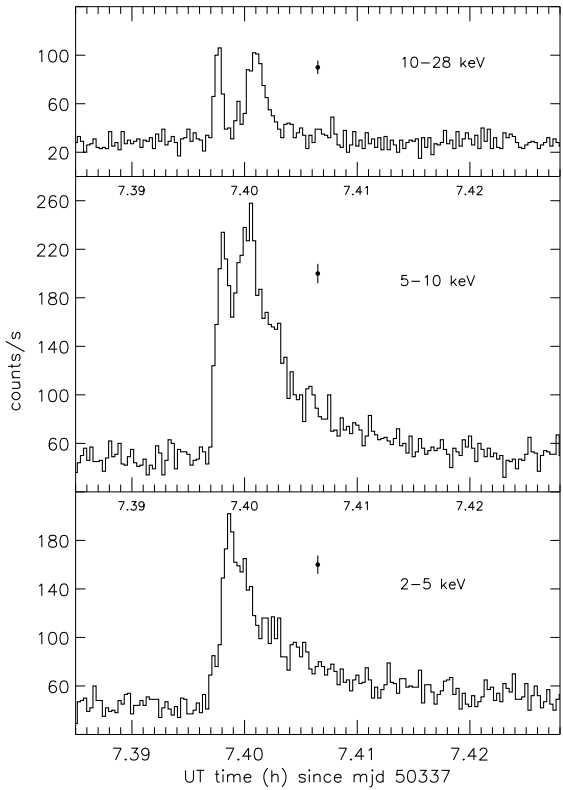,width=8.3cm,clip=t}
      \caption{
	       Time profiles of burst C in three energy bands.
	       The double-peaked structure becomes more evident at higher energies.
              }
         \label{Fig1}
   \end{figure}

\begin{table*}[hbt]
\caption{Summary of the characteristics of the observed bursts. Average spectral parameters are
         calculated for the first 30 s of burst data.
}
\protect\label{t:ae}
\begin{flushleft}
\begin{tabular}{lcccccccc}
\hline
\hline \noalign{\smallskip}
                                & burst A & burst B & burst C & burst D & burst E & burst F & burst G & burst H \\
\hline \noalign{\smallskip}
date                            & 1996 Aug.22 & 1996 Aug.29 & 1996 Sep.11 & 1997 Mar.12 & 
			          1997 Mar.30 & 1998 Sep.30 & 1998 Oct.7  & 1999 Sep.22 \\
UT (h)                          & 16.9158 & 5.6625 & 7.3972 & 5.3021 & 21.2650 & 21.9400 & 1.8097 & 7.7229 \\
\noalign{\smallskip}
$\tau_{2-28~{\rm keV}}$         & $15.0\pm 1.3$  & $14.5\pm 1.2$  & $15.1\pm 0.5$ & $14.1\pm 0.6$ & 
                                  $13.9\pm 1.0$  & $14.2\pm 1.2$  & $10.6\pm 1.1$ & $12.1\pm 1.0$ \\
$\tau_{2-8~{\rm keV}}$          & $16.8\pm 1.7$  & $18.0\pm 1.4$  & $20.1\pm 0.6$ & $16.4\pm 0.8$ &
			       	  $17.9\pm 1.6$  & $18.0\pm 1.7$  & $15.1\pm 1.8$ & $19.0\pm 2.1$ \\
$\tau_{8-28~{\rm keV}}$         & $ 7.3\pm 1.2$  & $ 5.8\pm 0.9$  & $ 6.9\pm 0.4$ & $ 6.3\pm 0.7$ & 
				  $ 4.1\pm 0.6$  & $ 4.4\pm 0.7$  & $ 6.8\pm 1.1$ & $ 6.1\pm 0.7$ \\
\noalign{\smallskip}
{\em k}T (keV)                  & $1.98^{+0.07}_{-0.06}$ & $1.85^{+0.04}_{-0.05}$ & $1.98^{+0.03}_{-0.02}$ & $1.96^{+0.03}_{-0.04}$ & 
				  $2.10\pm 0.06$         & $1.99^{+0.06}_{-0.05}$ & $2.26\pm 0.07$         & $2.33^{+0.08}_{-0.07}$ \\
${\rm R}_{bb}^{(a)}$            & $20.1^{+1.4}_{-1.3}$ & $21.8^{+1.3}_{-1.1}$ & $20.2^{+0.5}_{-0.4}$ & $20.9^{+0.8}_{-0.7}$ & 
				  $15.6\pm 0.9$        & $19.5^{+1.1}_{-1.0}$ & $15.0^{+1.0}_{-0.9}$ & $13.6^{+0.9}_{-0.8}$ \\
$\chi^{2}_{r}$                  & 1.57 & 0.89 & 1.53 & 1.13 & 1.69 & 1.36 & 1.18 & 0.75 \\
\noalign{\smallskip}
$I_{peak}^{(b)}$                & $4.69\pm 0.32$ & $4.21\pm 0.27$ & $4.36\pm 0.14$ & $4.87\pm 0.21$ & 
				  $4.66\pm 0.28$ & $4.81\pm 0.33$ & $4.32\pm 0.37$ & $4.68\pm 0.33$ \\
fluence$^{(c)}$
				& $2.91\pm 0.17$ & $2.81\pm 0.14$ & $3.14\pm 0.10$ & $2.93\pm 0.12$ &
				  $2.26\pm 0.12$ & $3.02\pm 0.16$ & $2.37\pm 0.15$ & $2.25\pm 0.13$ \\
\noalign{\smallskip}
\hline\noalign{\smallskip}
\noalign{$^{(a)}$ blackbody radius, for a 10 kpc distance; $^{(b)}$ peak intensity, in Crab units, 2-28 keV band; 
         $^{(c)}$ bolometric, in $10^{-6}~{\rm erg~cm}^{-2}$.  The bursts {\em e}-folding times $\tau$ are given in s.}
\end{tabular}
\end{flushleft}
\end{table*}

\section{Observations}

The Wide Field Cameras (WFC) on board the {\em BeppoSAX} satellite consist of two identical 
coded aperture multi-wire proportional counters (\cite{Jage97}) pointing in opposite directions.
Each camera covers a $40\degr \times 40\degr$ full width to zero-response field of view, the 
largest ever flown for an arcminute resolution X-ray telescope.
With their source location accuracy in the range $1\arcmin$--$3\arcmin$ (99\% confidence),
a time resolution of 0.488 ms, and an energy resolution of 18\% at 6 keV, the
WFCs are effective in studying X-ray transient phenomena in the
2--28 keV bandpass.  The imaging capability and the good instrument sensitivity
(5-10 mCrab on-axis in $10^{4}$ s, depending on the number of sources in the field) allow an accurate 
monitoring of complex sky regions, like the Galactic centre.

One of the main scientific objectives of the WFCs is the study of the timing and spectral 
behaviour of both transient and persistent sources of the Galactic Bulge region on time scales 
ranging from seconds to years.  
To this end, an observation program of systematic wide field monitoring of the Sgr~A 
sky region is being carried out since August 1996 (see e.g. \cite{Heis98,Heis99,Uber99}).
This program consists of a series of observations, each lasting $\sim 60$ ks, 
nearly weekly spaced throughout the two visibility periods (August-October and 
February-April) of the Galactic Centre region.

The WFC Galactic Bulge monitoring program is significantly contributing in the study of 
X-ray bursting sources. Up to now, a total of 15 new objects were discovered in $\sim 3.2$ years
observing time, thus enlarging the population of the bursters by $\sim 35\%$ (\cite{Heis99,Uber99}).
The data of the two cameras are systematically searched for bursts and flares by 
analyzing the time profiles of the detectors in the 2--11 keV 
energy range with a time resolution down to 1 s.  Reconstructed sky images are
generated for any statistically meaningful event and the accuracy of the reconstructed 
position, which of course depends on the burst intensity, is typically better than $5\arcmin$.
This analysis procedure has led to the identification of 
$\sim 950$ X-ray bursts (156 of which from the {\em Bursting Pulsar} GRO J1744$-$28) in 
a total of about $4\times10^{6}$ s net observing time (e.g. \cite{Cocc98a}).  

Whenever the WFCs point at the Galactic Centre region, 4U~1812$-$12 is in the field of view,
though at a rather offset position, being $\sim 18\degr$ away from Sgr~A ($l_{{\rm II}}=18.0\degr,
b_{{\rm II}}=2.4\degr$).
Due to the source's relatively low intensity and to the unfavourable pointing, the WFC data is not 
sensitive enough to study the persistent emission with sufficient accuracy.
A total of 8 X-ray bursts were detected at a position consistent with that of 4U~1812$-$12 
during all the time spent on the source by the WFCs both in primary and in secondary 
observing mode ($\sim 4$ Ms net time in 3.2 years).
None of the observed bursts can be associated with other known sources.

We analysed all the observed bursts, including the one already investigated by Burderi et al. (1997). 
The main characteristics of the observed bursts (hereafter burst A,B,...,H, chronologically) are 
summarised in Table 1.
One of the bursts, namely burst C, was observed with much better statistics than the
others, as it was in a more favourable position ($\sim 10\degr$ offset).  
For this reason, burst C was analysed with a higher time resolution than the others.

    \begin{figure}[hbt]
      \psfig{figure=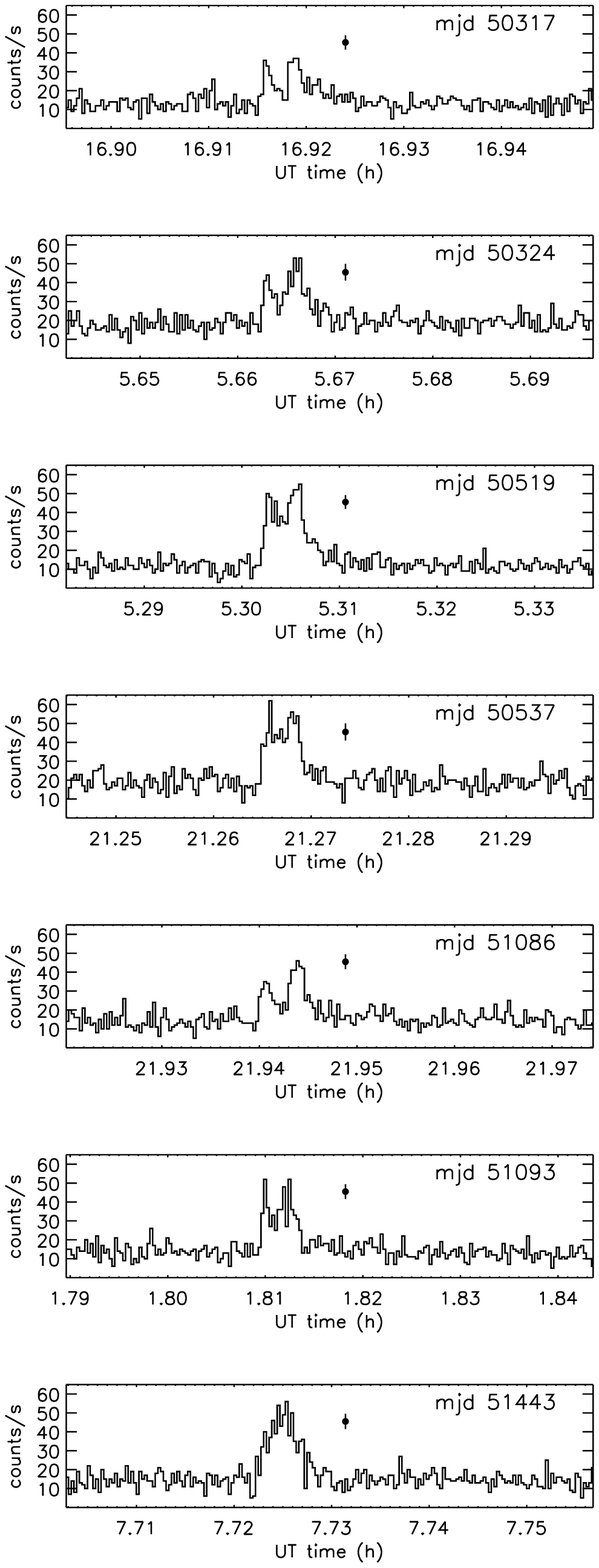,width=8.3cm,clip=t}
      \caption{
               Time histories of the A,B,D,E,F,G,H (starting from top) bursts in the
               8-28 keV energy band.  Eddington-luminosity effects are observed in all the
               profiles.
              }
      \label{Fig2}
   \end{figure}

\begin{table}[hbt]
\caption{ Time resolved spectral analysis of bursts A,B,D,E,F.
}
\protect\label{t:hc}
\begin{flushleft}
\begin{tabular}{llcl}
\hline
\hline\noalign{\smallskip} 
time range$^{(a)}$ & {\em k}T (keV) & $R_{\rm km}/d_{10~{\rm kpc}}$  & $\chi_{r}^{(b)}$ \\ 
\noalign{\smallskip} \hline\noalign{\smallskip} 
\multicolumn{4}{c}{{\bf mjd 50317 burst (A)}} \\
 $ 0s- 4s$      & $2.42^{+0.21}_{-0.18}$ & $15.2^{+2.8}_{-2.1}$ & $0.78^{(c)}$ \\
 $ 4s- 9s$      & $1.60 \pm 0.10$        & $35.3^{+5.1}_{-4.0}$ & $0.95^{(c)}$ \\
 $ 9s-14s$      & $2.36^{+0.14}_{-0.13}$ & $17.5^{+2.3}_{-1.8}$ & $1.22$ \\
 $14s-35s$      & $1.81^{+0.12}_{-0.10}$ & $18.0^{+2.6}_{-2.1}$ & $0.87$ \\
\noalign{\smallskip}
\multicolumn{4}{c}{{\bf mjd 50324 burst (B)}} \\
 $ 0s- 4s$      & $2.15 \pm 0.14$        & $16.7^{+2.5}_{-2.0}$ & $0.75$ \\
 $ 4s-10s$      & $1.54^{+0.07}_{-0.06}$ & $37.2^{+3.7}_{-3.1}$ & $0.47$ \\
 $10s-15s$      & $2.16^{+0.10}_{-0.09}$ & $20.2^{+2.0}_{-1.7}$ & $0.79$ \\
 $15s-35s$      & $1.76 \pm 0.14$        & $17.7^{+2.2}_{-1.8}$ & $0.76$ \\
\noalign{\smallskip}
\multicolumn{4}{c}{{\bf mjd 50519 burst (D)}} \\
 $ 0s- 4s$      & $2.68 \pm 0.16$        & $20.9^{+0.8}_{-0.7}$ & $1.68$ \\
 $ 4s-10s$      & $1.80^{+0.05}_{-0.04}$ & $30.7^{+1.7}_{-1.6}$ & $1.15$ \\
 $10s-15s$      & $2.24^{+0.08}_{-0.07}$ & $19.7^{+1.4}_{-1.2}$ & $1.39$ \\
 $15s-36s$      & $1.65 \pm 0.06$        & $20.8^{+1.6}_{-1.4}$ & $1.10$ \\
\noalign{\smallskip}                                                                                                                      
\multicolumn{4}{c}{{\bf mjd 50537 burst (E)}} \\
 $ 0s- 4s$      & $2.34 \pm 0.15$        & $15.6^{+2.2}_{-1.8}$ & $1.30$ \\
 $ 4s-10s$      & $2.15 \pm 0.08$        & $20.3^{+1.7}_{-1.5}$ & $1.39$ \\
 $10s-15s$      & $2.61^{+0.15}_{-0.14}$ & $13.0^{+1.6}_{-1.3}$ & $0.96$ \\
 $15s-32s$      & $1.33^{+0.10}_{-0.08}$ & $27.6^{+4.5}_{-3.5}$ & $1.09$ \\
\noalign{\smallskip}
\multicolumn{4}{c}{{\bf mjd 51086 burst (F)}} \\
 $ 0s- 5s$      & $2.48^{+0.18}_{-0.17}$ & $12.4^{+2.0}_{-1.6}$ & $0.72$ \\
 $ 5s-12s$      & $1.70 \pm 0.07$        & $31.7^{+3.0}_{-2.5}$ & $0.76$ \\
 $12s-17s$      & $2.51 \pm 0.12$        & $15.7^{+1.7}_{-1.4}$ & $0.84$ \\
 $17s-37s$      & $1.68 \pm 0.09$        & $18.9^{+2.5}_{-2.0}$ & $0.85$ \\
\noalign{\smallskip}
\multicolumn{4}{c}{{\bf mjd 51093 burst (G)}} \\
 $ 0s- 4s$      & $2.76 \pm 0.16$        & $11.7^{+1.6}_{-1.2}$ & $1.46$ \\
 $ 4s-10s$      & $2.10^{+0.09}_{-0.10}$ & $22.1^{+2.2}_{-1.9}$ & $0.57$ \\
 $10s-14s$      & $2.60^{+0.19}_{-0.17}$ & $12.9^{+2.0}_{-1.6}$ & $1.37$ \\
 $14s-34s$      & $1.91^{+0.14}_{-0.12}$ & $14.2^{+2.2}_{-1.8}$ & $1.23$ \\
\noalign{\smallskip}
\multicolumn{4}{c}{{\bf mjd 51443 burst (H)}} \\
 $ 0s- 4s$      & $2.40^{+0.16}_{-0.14}$ & $15.5^{+2.2}_{-1.8}$ & $0.81$ \\
 $ 4s- 9s$      & $2.50 \pm 0.10$        & $15.8^{+1.4}_{-1.3}$ & $1.35$ \\
 $ 9s-14s$      & $2.58 \pm 0.15$        & $13.6^{+1.8}_{-1.5}$ & $1.01$ \\
 $14s-34s$      & $1.54^{+0.16}_{-0.15}$ & $17.8^{+4.7}_{-3.1}$ & $0.78$ \\
\noalign{\smallskip} \hline\noalign{\smallskip}
\noalign{ $^{(a)}$ The zero is the burst time (see Table 1);
          $^{(b)}$ 27 d.o.f.; $^{(c)}$ 26 d.o.f.}
\end{tabular}
\end{flushleft}
\end{table}

\section{Data Analysis and Results}

Energy-resolved time histories of the bursts were constructed by accumulating 
the detector counts associated with the shadowgram obtained for the 
sky position of the analysed source, thus improving the signal-to-noise ratio of the profile.  
For a given source, the background is the sum of (part of) the diffuse X-ray background, the particles 
background and the contamination of other sources in the field of view.
Source contamination is the dominating background component for crowded sky fields like 
the Galactic Bulge.  Nevertheless, the probability of source confusion during a short 
time-scale event (10--100 s) like an X-ray burst is negligible.
The time profiles of burst C were accumulated in three different energy bands 
(2--5, 5--10, and 10--28 keV, see Fig. 1). The time histories of the other bursts, due 
to their limited counting statistics, were obtained for the bands 2-8 and 8-28 keV only (Fig.2).
The 2-28 keV time profiles of all the bursts are characterized by fast rise times (within a few seconds) 
and longer exponential decay with {\em e}-folding times of $\sim 15$ s.  
The high energy time histories of almost all the bursts show clear evidence for double-peaked profiles 
(see Fig.1 and Fig.2) and their {\em e}-folding times are significantly shorter ($\sim 4-7$ s) 
than the ones of the low energy profiles ($\sim 15-20$ s), as reported in Table 1.

The integrated spectra of the eight bursts are all consistent with absorbed blackbody radiation with 
average colour temperatures of $\sim 2$ keV and an average blackbody radius of the emitting sphere of 
$\sim 20$ km assuming a standard 10 kpc source distance (Table 1).  The spectra were subtracted for the 
source persistent emission, which accounts for only $\sim 0.5\%$ of the burst peak intensity.
The $N_{\rm H}$ parameter could not be satisfactorily constrained for any of the bursts, so we kept 
its value fixed according to the interpolated value computed at the source position, namely
$N_{\rm H} = 7.3\times 10^{21}~{\rm cm}^{-2}$ (\cite{Dick90}).

Time-resolved spectra were accumulated for all the bursts, in order to study the time
evolution of their spectral parameters.  Thanks to the good counting statistics, the spectral 
analysis of burst C could be performed with a time resolution of 1 s. 
Conversely, for each of the other bursts only four time resolved spectra could be obtained with a 
poorer time resolution (4 s). The time intervals of the four spectra were chosen to match  
the first-peak, interpeak, second-peak and decay phases in the corresponding 8-28 keV burst time 
profiles (Fig.2). 
Since all the obtained spectra are consistent with absorbed blackbody emission, the time 
histories of the colour temperature and the emitting sphere radius can be determined 
(Table 2 and Fig.3).
The blackbody radii were calculated assuming isotropic emission at a source distance of 10 kpc and 
not correcting for gravitational redshift and conversion to effective blackbody temperature from color 
temperature (see \cite{Lewi93} for details). 
A radius expansion by a factor of $\sim 5$ is observed in burst C while, probably due to the larger time 
bins used, a lower expansion factor ($\sim 2$) is obtained for all the other bursts but burst H,
which does not show evidence for radius variations.

    \begin{figure}[htb]
      \psfig{figure=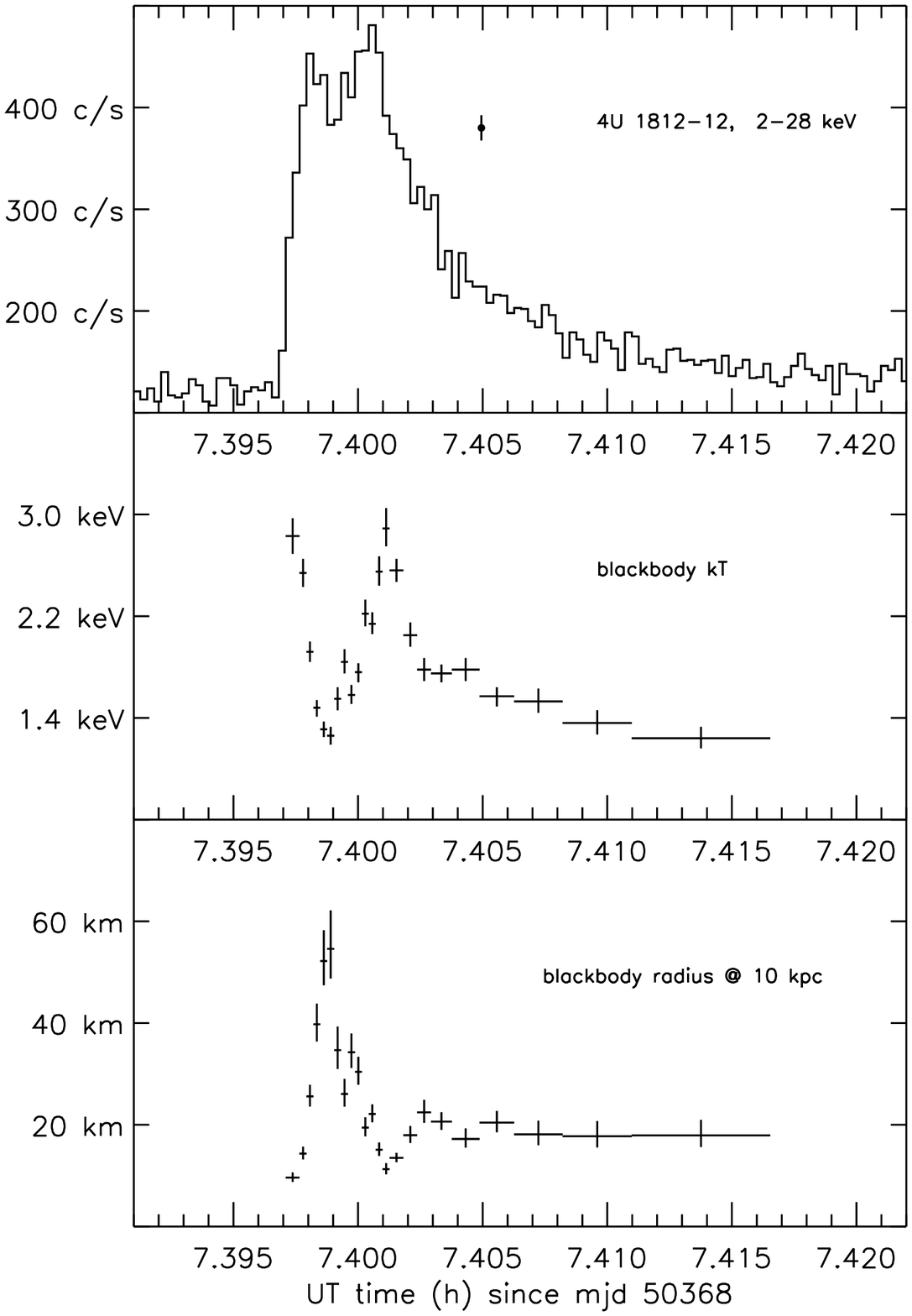,width=8.3cm,clip=t}
      \caption{
	       Time histories of the 2-28 keV count rate (upper panel),
	       of the blackbody colour temperature (central panel), and
	       of the blackbody radius (lower panel), as observed for
	       burst C. 
              }
         \label{Fig3}
   \end{figure}

\section{Discussion}

Following the classification proposed by \cite{Hoff78}, X-ray bursts are classified in two 
main types ({\em type-I}, {\em type-II}, see Lewin, van Paradijs \& Taam (1993) for a 
comprehensive review).

On the basis of the spectral and timing properties of the eight bursts observed by the WFCs,
it is apparent that 4U~1812$-$12 is a type-I burster. 
In fact, the blackbody emission and the measured colour temperatures of $\sim 2$ keV are consistent 
with type-I bursting. Moreover, spectral softening is observed in the time resolved spectra of the 
bursts (Table 2 and Figure 3), and the bursts time profiles can be fitted with exponential decays 
whose characteristic times are significantly shorter at higher energies (see e.g. Table 1). 
These results confirm the measurements obtained 15 years earlier by {\em Hakucho} (\cite{Mura83}), 
indicating 4U~1812$-$12 is a neutron-star low-mass X-ray binary. 

The photospheric radius expansion derived from the time resolved spectral analysis of most of the  
observed bursts can be interpreted as adiabatic expansion during a high luminosity (Eddington-limit) 
type-I burst.  Actually, the double-peaked profiles observed in the high energy (above 8 keV) time 
histories of the bursts (with the only exception of burst H) are typical of super-Eddington events 
(e.g. \cite{Lewi95}).  
Even though burst H is not double-peaked, its peak luminosity is consistent with those of the other 
observed events. Moreover, its 8--28 keV time profile could be associated with the flat-top profiles 
of some observed Eddington-limited type-I bursts (\cite{Lewi95}), so burst H too can be regarded as 
an event with peak luminosity close to the Eddington limit.  
We also notice that the two bursts which show less clear evidence for photospheric radius expansion,  
namely bursts E and H, are the less energetic ones, their fluences being the lowest observed (Table 1).
We can regard their total energy release ($\sim 5\times 10^{39}$ erg assuming a 4 kpc distance, see later) 
as the minimum needed to drive the expansion of the photosphere of the neutron star in 4U~1812$-$12. 

Eddington-luminosity X-ray bursts can lead to an estimate of the source distance, assuming the burst 
emission to be isotropic and the peak flux to be very close to the Eddington luminosity.
Actually, the peak intensities observed for the eight events are all consistent with a constant 
value of $4.53 \pm 0.09$ Crab (2--28 keV); the associated reduced $\chi^2$ is 1.03 
for 7 d.o.f..
This average peak intensity extrapolates to an unabsorbed bolometric flux of 
$(15.03 \pm 0.29)\times 10^{-8}~{\rm erg~cm}^{-2}~{\rm s}^{-1}$.  
The consistency of the peak luminosities of all the bursts with a constant value supports the 
adoption of the peak bolometric intensities of super-Eddington bursts as a standard candle.
An average luminosity of Eddington-limited bursts was empirically calculated by Lewin, van Paradijs and 
Taam (1995) on a sample of bursters whose distance was estimated with other methods: 
a luminosity value of ($3.0\pm 0.6\times 10^{38}~{\rm erg~s}^{-1}$) was obtained.
The adoption of this standard luminosity leads to a distance value $d = 4.1\pm 0.5$ kpc for 4U~1812$-$12. 
On the other hand, assuming the theoretical Eddington luminosity for a typical  
$1.4~{\rm M}_{\sun}$ neutron star ($\sim 2\times 10^{38}~{\rm erg~s}^{-1}$) we obtain $d \sim 3.3$ kpc.
For the calculated distance of $\sim 4$ kpc, and with the simple assumptions on the burst emission made 
in Section 3, an average radius of $8\pm 1$ km for the blackbody emitting region during the bursts is 
obtained. 
This value, which supports the neutron-star nature of the collapsed object, should indeed be regarded as 
a lower limit for the actual neutron star radius, according to Ebisuzaki, Hanawa \& Sugimoto (1984).

Assuming the source's persistent spectrum to be consistent with the one suggested by Barret 
et al. (2000) for the X-ray bursters in low state, i.e. a Comptonized spectrum with electron temperature 
$k{\rm T}_{e} \sim 25$~keV and $\tau \sim 3$, the bolometric luminosity of 4U~1812$-$12 can be 
extrapolated.
For a distance of $\sim 4$ kpc and an average 2--10 keV persistent intensity of 
$\sim 4\times10^{-10}~{\rm erg~cm}^{-2}{\rm s}^{-1}$, we obtain 
$L_{bol} \sim 5.6\times 10^{36}~{\rm erg~s}^{-1}$.  
We also derive, for a canonical $1.4~{\rm M}_{\odot}$ neutron star with a radius of 10 km, an average 
accretion rate of $\sim 5\times10^{-10}~{\rm M}_{\odot}{\rm y}^{-1}$.
These values are common among low-mass X-ray binaries.

Due to the non-continuous WFCs coverage of the Galactic Centre region and to the {\em BeppoSAX} 
orbit characteristics, we cannot accurately establish the burst occurrence rate of 4U~1812$-$12. 
Anyway the minimum observed intervals are 6.53 d and 6.16 d for bursts A-B (August 1996) and F-G 
(October 1998) respectively, and such intervals are of the same order of magnitude than the one 
measured by {\em Hakucho} in 1982 (4.61 d).  Under the hypothesis that the above values are the 
actual wait times for bursts B and G, we can calculate the ratio $\alpha = E_{p}/E_{b}$, where 
$E_{b}$ and $E_{p}$ are the bolometric fluences of the burst and of the persistent emission between 
two contiguous bursts, respectively. 
Average 2--10 keV intensities of $(4.4\pm 0.4) \times 10^{-10}~{\rm erg~cm}^{-2}{\rm s}^{-1}$ and
$(3.5 \pm 0.9) \times 10^{-10}~{\rm erg~cm}^{-2}{\rm s}^{-1}$ were measured by {\em RXTE}-ASM 
between bursts A-B and F-G respectively. Again, with the former assumptions on the spectrum of the 
persistent emission, we determine $\alpha = (6.4\pm 1.0)\times 10^{2}$ and 
$\alpha = (5.8\pm 1.9)\times 10^{2}$ for the two events.
These values are consistent with each other and are within the observed range ($10-10^{3}$, 
distribution peaking at $\sim 10^{2}$) for the $\alpha$ parameter of known X-ray bursters, even if 
on the higher side. This is suggestive of helium-burning with no spare fuel left for the next burst, 
and possibly of steady burning of part of the accreted matter ({\cite{Lewi93, vanP88}).  

As pointed out above, the eight bursts we analysed show very similar features. Moreover, 
their characteristics are also consistent with those of the bursts detected by {\em Hakucho} 
(\cite{Mura83}). For the {\em Hakucho} bursts, 1--22 keV peak intensities of $\sim 4.6$ Crab were 
measured.
Also the event Burderi et al. (1997) reported to have very different burst parameters, i.e. single 
peak profile and much lower luminosity (1.2 Crab in 1.5--10 keV), is actually very similar to the 
others, according to the results of our re-analysis (burst D).
The above consistencies, together with the $\sim 5-6$ d burst wait times observed by both {\em Hakucho}
and {\em BeppoSAX}, suggest the burst characteristics of the binary 4U~1812$-$12 to be remarkably stable 
in observations spanning $\sim 15$ years apart.

\begin{acknowledgements}
  We thank the staff of the {\em BeppoSAX Science Operation Centre} and {\em Science
  Data Centre} for their help in carrying out and processing the WFC Galactic Centre
  observations. The {\em BeppoSAX} satellite is a joint Italian and Dutch program.
  M.C., A.B., L.N. and P.U. thank Agenzia Spaziale Nazionale ({\em ASI}) for grant support.
  M.C. also thanks M. Federici (IAS) for technical support.
\end{acknowledgements}

\end{document}